\documentclass[twocolumn]{aastex701}

\begin{document}

\title{Overestimated Pressure Broadening Misleads Model Spectra in Cool M Dwarf Stars} 

\correspondingauthor{Ana Glidden}
\email{aglidden@mit.edu}

\author[0000-0002-5322-2315]{Ana Glidden}
\affiliation{Department of Earth, Atmospheric and Planetary Sciences, Massachusetts Institute of Technology, Cambridge, MA 02139, USA}
\affiliation{Kavli Institute for Astrophysics and Space Research, Massachusetts Institute of Technology, Cambridge, MA 02139, USA}
\email{aglidden@mit.edu}

\author[0000-0002-0929-1612]{Veronika Witzke}
\affiliation{Institute of Physics, University of Graz, A-8010 Graz, Austria}
\email{veronika.witzke@uni-graz.at}

\author[0000-0002-8842-5403]{Alexander I. Shapiro}
\affiliation{Institute of Physics, University of Graz, A-8010 Graz, Austria}
\affiliation{Max-Planck-Institut f\"ur Sonnensystemforschung, Justus-von-Liebig-Weg 3, 37077 G\"ottingen, Germany}
\email{alexander.shapiro@uni-graz.at}

\author[0000-0002-6892-6948]{Sara Seager}
\affiliation{Department of Physics and Kavli Institute for Astrophysics and Space Research, Massachusetts Institute of Technology, Cambridge, MA 02139, USA}
\affiliation{Department of Earth, Atmospheric and Planetary Sciences, Massachusetts Institute of Technology, Cambridge, MA 02139, USA}
\affiliation{Department of Aeronautics and Astronautics, Massachusetts Institute of Technology, 77 Massachusetts Avenue, Cambridge, MA 02139, USA}
\email{seager@mit.edu}

\begin{abstract}

Available one-dimensional stellar models fail to reproduce the observed spectrum of the ultracool M dwarf TRAPPIST-1. In particular, current models predict strong iron hydride (FeH) absorption due to the Wing–Ford bands at 0.99~\micron, yet this spectral feature is only weakly present in TRAPPIST-1 and other mid-to-late M dwarf stars. Additionally, the shape of the continuum between the water bands in the near-infrared does not match between models and observations. Here, we show that assumptions about pressure broadening, specifically van der Waals broadening, have a dramatic effect on modeled broadband spectral features. We use {\tt Merged Parallelized Simplified-ATLAS} to generate synthetic spectra over a range of van der Waals broadening strengths, adopting 1D {\tt PHOENIX} temperature–pressure structures. We find that minimal broadening best matches the observed FeH profile at 0.99~\micron~and in the pseudocontinuum between the large water bands. These results suggest that broadening prescriptions derived for Sun-like stars are not valid for lower-mass stars and that pressure broadening for molecular lines in cool stellar atmospheres must be reevaluated. Refining pressure broadening treatments will improve the accuracy of M dwarf spectral models, enabling more reliable determinations of stellar properties and atmospheric compositions of planets orbiting M dwarfs. 

\end{abstract}

\section{Introduction} \label{sec:intro}

M dwarfs are the most common stars in our Galaxy, yet they are still poorly understood. Despite extensive theoretical modeling work to match their observed spectra, discrepancies still remain \citep[e.g.,][]{Husser2013, Iyer2023, Lim2023, Hauschildt2025, Piaulet-Ghorayeb2025}. Many of these differences stem from historical biases toward studying Sun-like stars. The accuracy of M dwarf spectral models has come under renewed scrutiny with the advent of JWST. With JWST, we can finally probe the atmospheres of terrestrial planets for the first time---places where life could arise. However, whether viewed through transmission or emission, inaccurate stellar models limit planetary atmospheric constraints and induce unknown uncertainties on inferred planetary properties \citep[e.g.,][]{Rackham2018, deWit2024, Fauchez2025}. Thus, accurate stellar models are needed to not only better understand the underlying physics of stars, but to robustly unravel exoplanet properties. 

M dwarf spectra are particularly challenging to model because they are  dominated by millions of molecular lines both in the optical and infrared. Molecular opacities are poorly constrained and thus limit the accuracy of stellar models. Metal hydrides like TiO, VO, and FeH have particularly complex spectra that are less well understood \citep{WingFord1969, Hargreaves2010}. M dwarfs are also highly active \citep[e.g.,][]{HawleyPettersen1991, Hawley2014, Schmidt2014}. The line shapes are impacted by both granulation and magnetic activity, which vary with time. Magnetic activity also leads to unpredictable, inhomogeneous changes to the stellar surface such as those due to starspots, faculae, flares, and granulation. To further complicate spectra, some have proposed M dwarf atmospheres may be cool enough for molecules to condense into clouds, such as MgSiO$_3$, Mg$_2$SiO$_4$, Fe, and Al$_2$O$_3$ clouds \citep{Tsuji1996, Iyer2023, Xuan2024}. 

Thus, state-of-the-art stellar modeling codes have faced many challenges as they push to cool M dwarf stars. {\tt PHOENIX} \citep{Hauschildt1997} is a non-local thermodynamic equilibrium 1D code widely used in the astronomical community, which has been frequently updated since its release. Recently, an updated version of {\tt PHOENIX} called the {\tt NewEra} model grid was released \citep{Hauschildt2025}. The {\tt NewEra} uses a larger and more precise database of atomic and molecular lines. Here, we use our in-house stellar synthesis code, {\tt Merged Parallelized Simplified} {\tt (MPS)}-{\tt ATLAS} to test assumptions about the input opacity sources \citep{Witzke2021}. {\tt MPS-ATLAS} produces stellar atmospheric models quickly through merged parallelization. The radiative transfer models incorporate key stellar parameters, such as effective temperature (T$_{\text{eff}}$), surface gravity (log($g$)), metallicity, and alpha enhancement. {\tt MPS-ATLAS} models the stellar photosphere, where the photons we observe are emitted from. {\tt MPS-ATLAS} is built on the heritage code {\tt ATLAS9} \citep{Kurucz2017} and has been validated against the {\tt ATLAS9} and {\tt PHOENIX} grids.

Recent observations show a mismatch between modeled spectra and observations \citep{Lim2023, Piaulet-Ghorayeb2025}. \citet{Lim2023, Piaulet-Ghorayeb2025} invoke a low log($g$) for their {\tt PHOENIX} models to match the TRAPPIST-1 data, which is physically incompatible with TRAPPIST-1's dwarf status. The main effect from decreasing log($g$) on emergent spectrum comes from a decrease in pressure and thereby pressure broadening. Surface gravity is well constrained for TRAPPIST-1, thus motivating us to find an alternative explanation for the mismatch between data and models. Here, we investigate if better agreement can be found by directly reducing pressure broadening, which, in contrast to log($g$), is poorly constrained.

Generally, pressure broadening is caused by collisional interactions between particles, such that their atomic energy levels are perturbed, leading to changes in their observed spectrum \citep{Gray2022}. One type of pressure broadening is van der Waals broadening, which is caused specifically by neutral perturbers such as H I, He I, and H$_2$. Van der Waals broadening impacts most lines, especially for cool stars. Van der Waals broadening has not been measured at the temperatures and pressures corresponding to the conditions of M dwarf atmospheres. Theoretical estimates are also very limited. Consequently, the treatment of van der Waals broadening in existing radiative transfer codes is pretty rudimentary.  The broadening parameters are often empirically derived from observations of solar spectral lines or calculated using approximations developed for atomic collisions (e.g., the Uns{\"o}ld approximation). 

Our Letter is structured as follows. In Section~\ref{sec:methods}, we summarize how we model the stellar spectrum and compare our results with observed spectra of M dwarf TRAPPIST-1. Then, in Section~\ref{sec:results} we outline the results of our analysis. In Section~\ref{sec:summary}, we summarize our findings and their limitations and outline the path forward.\\


\section{Methods} \label{sec:methods}

We aim to compare synthetic spectra with different pressure broadening strengths to observed spectra in order to evaluate the impact of van der Waals broadening assumptions. We use the {\tt MPS-ATLAS} radiative transfer code to synthesize opacity and spectra as it allows flexibility to generate our own high-resolution opacity tables with different broadening values. For consistency with previous work \citep{Lim2023, Piaulet-Ghorayeb2025}, we use the 1D temperature-pressure profiles from the {\tt PHOENIX} grid of models \citep{Hauschildt1997, Husser2013} \footnote{https://phoenix.astro.physik.uni-goettingen.de} as input.

We calculate opacity tables with a resolving power of $\text{R}= 200,000$ in the spectral range of $0.5~\mu m$ to $2.5~\mu m$ using {\tt MPS-ATLAS}. For our opacity calculations, we use atomic lines from the VALD3 database \citep{Ryabchikova2015}, H$_2$O \citep{Partridge1997}, TiO isotopologues \citep{Schwenke1998}, FeH \citep{Dulick2003}, and additional molecules (H$_2$, CH, NH, C$_2$, CN, CO, MgH, SiH and SiO) compiled by R.~Kurucz\footnote{http://kurucz.harvard.edu/linelists/linescd/}.
The pressure and temperature dependence of the van der Waals broadening was calculated according to the procedure described in \citet{Castelli_2005}. The arbitrary broadening parameter is taken from Kurucz's line lists. 
Since the estimated metallicity of TRAPPIST-1 is $\mathrm{Fe /H} = 0.04 \pm 0.08$ \citep{Gillon2016, VanGrootel2018}, which is very close to solar metallicity, we used the solar composition \citep{Asplund2006} in all our calculations. For all included molecules, we calculate a series of opacity tables where we scaled the arbitrary van der Waals broadening parameter with a factor of 0, 0.01, 0.05, 0.1, 0.5, and 1.0.

We use {\tt MPS-ATLAS} to calculate the emergent flux over a range of 11 viewing angles, from $\mu = 1.0$ (disk center) down to $\mu = 0.05$ (limb). We then integrate the fluxes to calculate the total flux across the surface of the disk. 

To compare with our synthetic spectra, we use published data of TRAPPIST-1 \citep[M8V; T$_{eff}$=$2566\pm26$; log($g$)=$5.2396^{+0.0056}_{-0.0073}$;][]{Agol2021} taken with the JWST Near-Infrared Imager and Slitless Spectrograph \citep[NIRISS;][]{Doyon2023} in the Single-Object Slitless Spectroscopy \citep[SOSS;][]{Albert2023} mode as described in \citet{Lim2023} as our fiducial dataset. NIRISS/SOSS spans $0.6-2.8\mu$m and has a resolution of R$\simeq700$ around the FeH feature ($\sim$1$\mu$m) and H$\alpha$ (0.656 $\mu$m). 

We bin the spectra using the wrapper of {\tt SpectRes} \citep{Carnall2017} found in {\tt POSEIDON} \citep{MacDonald2023}. {\tt SpectRes} accurately resamples spectral data while conserving integrated flux. We bin to R=120 and normalize our modeled spectra to the peak of the spectrum without van der Waals broadening.


\section{Results} \label{sec:results}

We describe our results in context with stellar atmosphere models before presenting our main findings. 

\begin{figure*}[!ht]
    \centering
    \includegraphics[width=\textwidth]{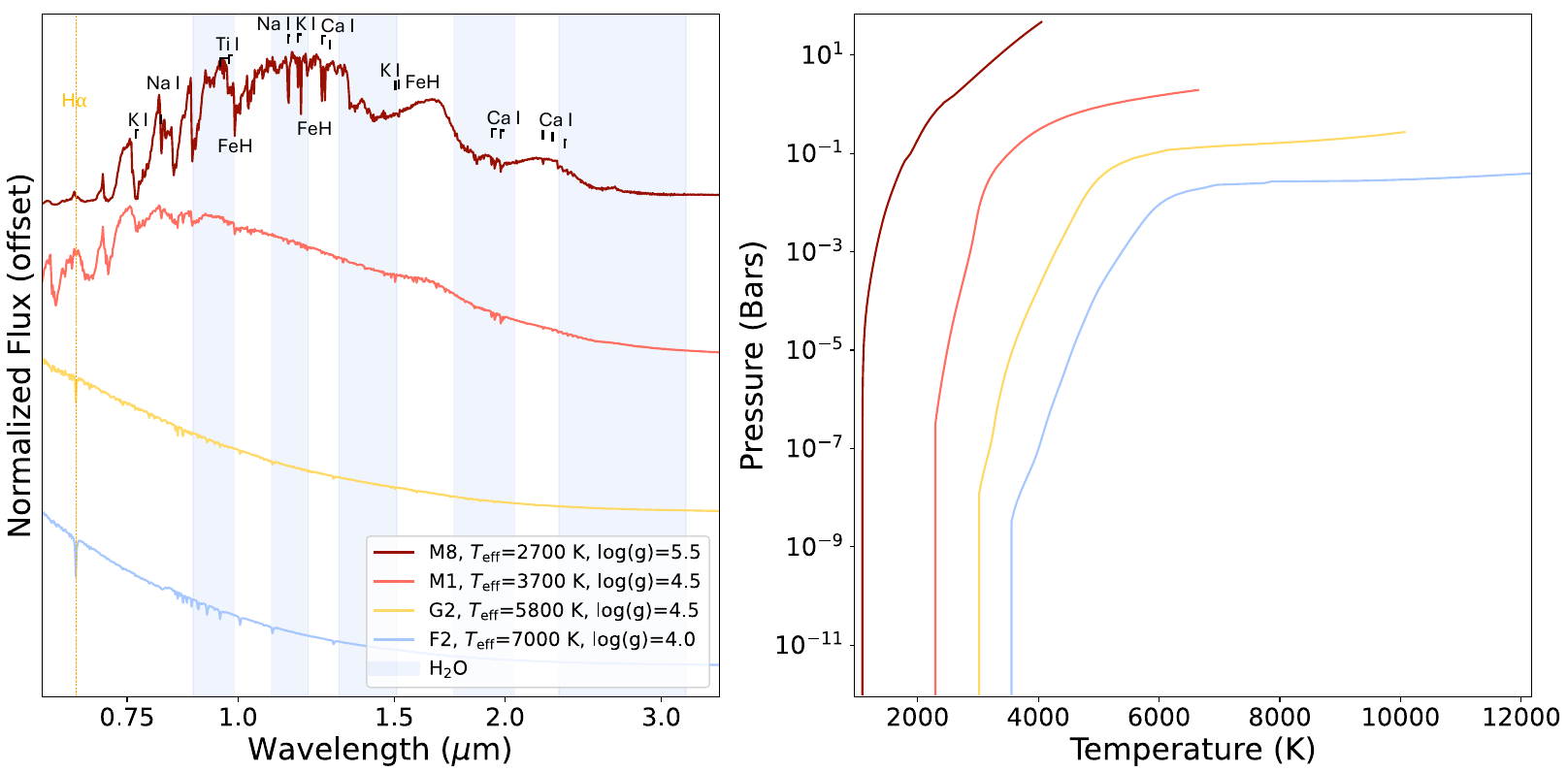}
    \caption{\textbf{Spectra of different stellar types.} On the left, scaled and vertically shifted flux from {\tt NewEra} stellar grid is shown on the y-axis and wavelength is shown in microns on the x-axis. On the right, the corresponding pressure-vs-temperature profiles are shown, with temperature in Kelvin on the x-axis and pressure in bars on the y-axis. Prominent atomic and molecular features are annotated \citep[][and references therein]{Cushing2005, Rajpurohit2018}. Molecular features increase as we move toward cooler stars. In particular, the FeH and water features (shaded blue bands) stand out in the M8 model and are not present in hotter stars.}
    \label{fig:stellartypes}
\end{figure*}

\subsection{M Dwarf Model Spectra Complexity}
M dwarfs are particularly difficult to model as they are cool enough for the formation of numerous molecules, which imprint a forest of molecular features. The M8 dwarf shown in Figure \ref{fig:stellartypes} is peppered with molecular absorption lines from oxygen-bearing molecules (e.g., TiO, CaH, VO, and CO); metal hydrides (AlH, CaH, CrH, FeH, MgH, TiH), and atomic lines (Na I, K I, Cs I, Ca I, etc.). In contrast, for the hotter F2 star, only atomic hydrogen lines are clearly visible. Molecular lines are particularly challenging due to their sheer number and poorly constrained broadening parameters for temperatures and pressures typical of M dwarf atmospheres. In particular, the broadening coefficients for temperature dependence is not known to high accuracy. Metal hydrides like FeH are particularly challenging to determine \citep{Dulick2003, Hargreaves2010}.

\begin{figure*}[!ht]
    \centering
    \includegraphics[width=0.85\textwidth]{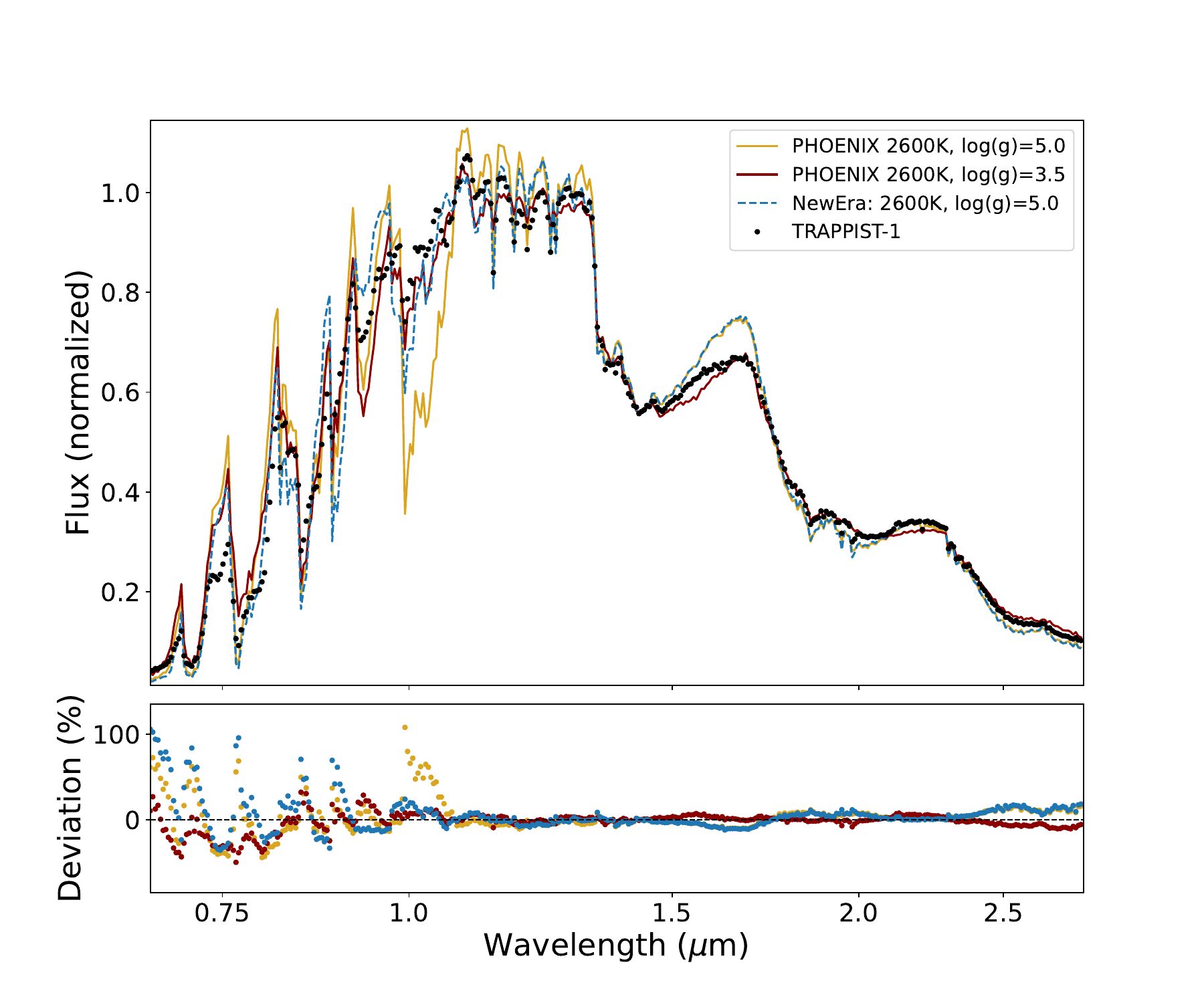}
    \caption{\textbf{TRAPPIST-1 observations compared with {\tt PHOENIX} and {\tt NewEra} stellar grids.} In the upper panel, the scaled flux is shown on the y-axis and wavelength is shown in microns on the x-axis. Two {\tt PHOENIX} spectra are shown for a star of 2600~K with log($g$) of 5.0 (gold) and 3.5 (magenta), respectively. The {\tt NewEra} grid spectra is shown (dashed blue) for a 2600~K with log($g$) of 5.0 star. NIRISS/SOSS data for TRAPPIST-1 are shown for comparison (black dots). The deviation between each synthetic spectra and the data is shown as a percentage in the lower panel. Disagreement between the synthesized and observed spectra is most pronounced around the FeH feature at $0.99\mu$m and between the large water features.  Decreasing log($g$) significantly improves the fit to the FeH feature for the {\tt PHOENIX} grid. However, a surface gravity of log($g$)=3.5 is consistent with a subgiant star, not an M dwarf. Observations show that TRAPPIST-1 is a main-sequence star with log($g$)$\approx5.2$ \citep{Agol2021}. This discrepancy points to the need to improve current 1D models. With {\tt NewEra}, we can now better match stellar data with reasonable stellar parameters. However, while {\tt NewEra} is more consistent with the FeH feature, disagreement persists in the pseudocontinuum between the large water bands centered around 1.4~\micron~and 1.9~\micron.}
    \label{fig:modelvdata}
\end{figure*}

A large point of tension between models and data for TRAPPIST-1 is the FeH feature \citep[Wing-Ford band,][]{WingFord1969} at $0.99~\mu$m. To illustrate the challenge with FeH, we compare three synthesized stellar spectra with JWST NIRISS/SOSS data of TRAPPIST-1 (Figure \ref{fig:modelvdata}). In order to accurately capture the spectral feature, we must assume an unphysically low surface gravity. In contrast, the improved {\tt NewEra} grid \citep{Hauschildt2025}, which has updated opacities, but consistent stellar models, allows for a better fit to the FeH feature with the correct stellar temperature and surface gravity. However, the  {\tt NewEra} spectral model is still a poor fit for the continuum between the large water features around $=1.7$~\micron~and beyond. While {\tt NewEra} is an improvement over {\tt PHOENIX}, this discrepancy speaks to the need for additional improvement in stellar spectral modeling.

\subsection{FeH Feature is Best Fit with Minimized Van der Waals Broadening}

\begin{figure*}[!ht]
    \centering
    \includegraphics[width=0.85\textwidth]{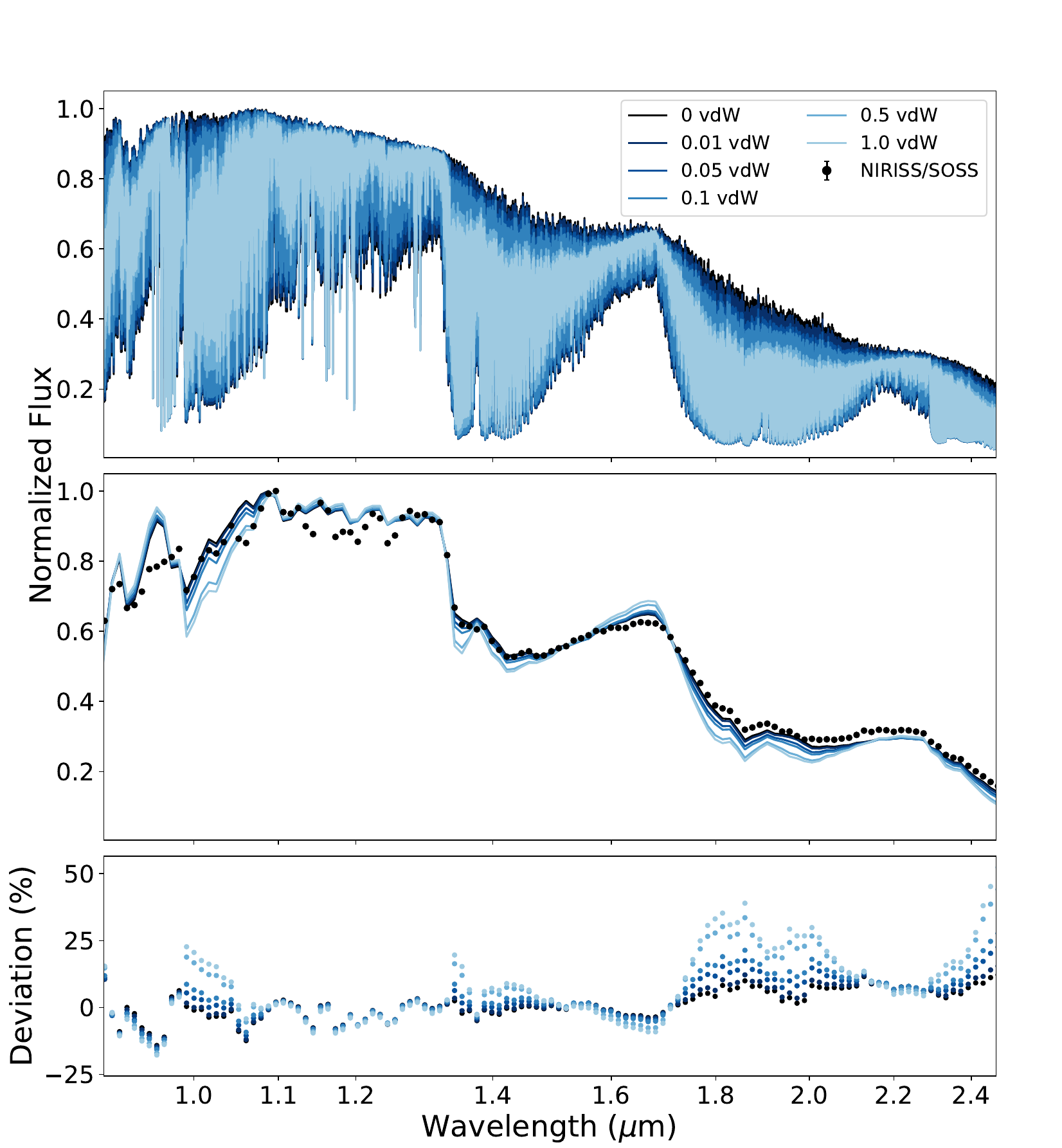}
    \caption{\textbf{Comparison of spectra synthesized using a range of van der Waals (vdW) broadening strengths with observed TRAPPIST-1 data.} For the upper two panels, the upper y-axis shows the normalized flux, while the x-axis shows the wavelength in microns. Modeled spectra are shown with lines at R$=$200,000 (top) and compared with the data at R=120 (middle). We use 1D {\tt PHOENIX} atmospheric structure (pressure--temperature profiles) for a star of 2600 K and log($g$)=5.0 as input to {\tt MPS-ATLAS}. The lowest y-axis shows the percentage deviation between the data and synthesized spectra at R=120. Reducing van der Waals broadening leads to better agreement, though not a complete match between the calculations and observations.}
    \label{fig:vdw}
\end{figure*}

Minimizing van der Waals broadening gives the best fit to the FeH feature in the TRAPPIST-1 data. To demonstrate this, we compare the JWST NIRISS/SOSS data of TRAPPIST-1 \citep{Lim2023} with a series of synthesized spectra using different scaling factors for van der Waals broadening, but the same underlying atmospheric structure. The fit is best for the spectra obtained with minimized van der Waals broadening (Figure \ref{fig:vdw}).

\begin{figure*}[!ht]
    \centering
    \includegraphics[width=0.8\textwidth]{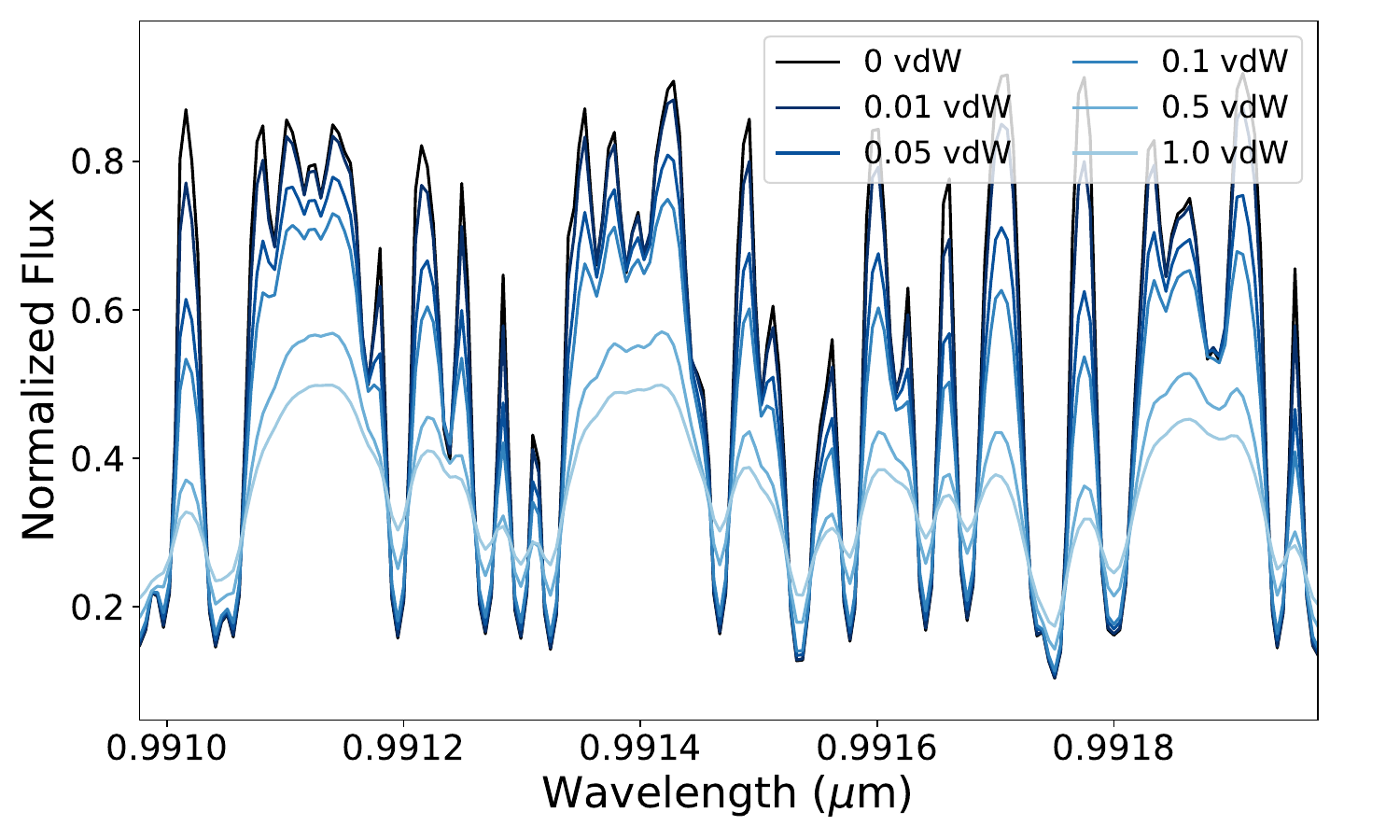}
    \caption{\textbf{Effect of van der Waals broadening on individual spectral lines.} The normalized flux is shown on the y-axis and the wavelength is shown in microns on the x-axis. The models are shown at R=200,000 and shown within the FeH Wing-Ford band. The lines are colored according to the fractional amount of van der Waals (vdW) broadening with increased broadening shown in lighter colors.}
    \label{fig:vdw_highres}
\end{figure*}

Pressure broadening imprints a very strong effect on {\it broadband} molecular features---and the shape of TRAPPIST-1's spectrum---and can be understood by recalling the effect of pressure broadening on strong lines. Strong lines are saturated, such that increased collisional broadening increases the width of saturated lines without affecting their depth. The combined effect of this increase in width in millions of lines makes the strength of the broadband features highly dependent on pressure broadening. A useful analogy is to consider individual molecular lines as shades blocking the escape of photons. When the lines are narrow, photons can still escape through opacity ``windows” between the lines. Broadening of the lines closes these opacity windows and dramatically reduces the amount of escaping photons. To demonstrate the impact, we show individual spectral lines at R$=$200,000 across the range of broadening strengths within the FeH Wing-Ford band (Figure \ref{fig:vdw_highres}).

\subsection{Peak between Broad Water Band Features also Best Fit with Minimized Van der Waals Broadening}

Minimizing van der Waals broadening also harmonizes disagreement between models and data around the water features. Water is the most significant source of opacity in M dwarf stars above 1~$\micron$. Reliable modeling of the water features is widely acknowledged as a persistent source of uncertainty. In particular, there are broad water bands centered around 1.4 and 1.9~$\micron$. Between these bands, there is a peak in the pseudocontinuum at $\sim1.7~\micron$. There is significant tension between the models and data over the shape of the peak. Decreasing van der Waals broadening flattens the slope of the peak, leading to better agreement with observations overall, but particularly around the large water feature around 1.8 to 2.1$~\micron$.

\subsection{Comparison of Broadening Implementations}

\begin{figure*}[!ht]
    \centering
    \includegraphics[width=0.8\textwidth]{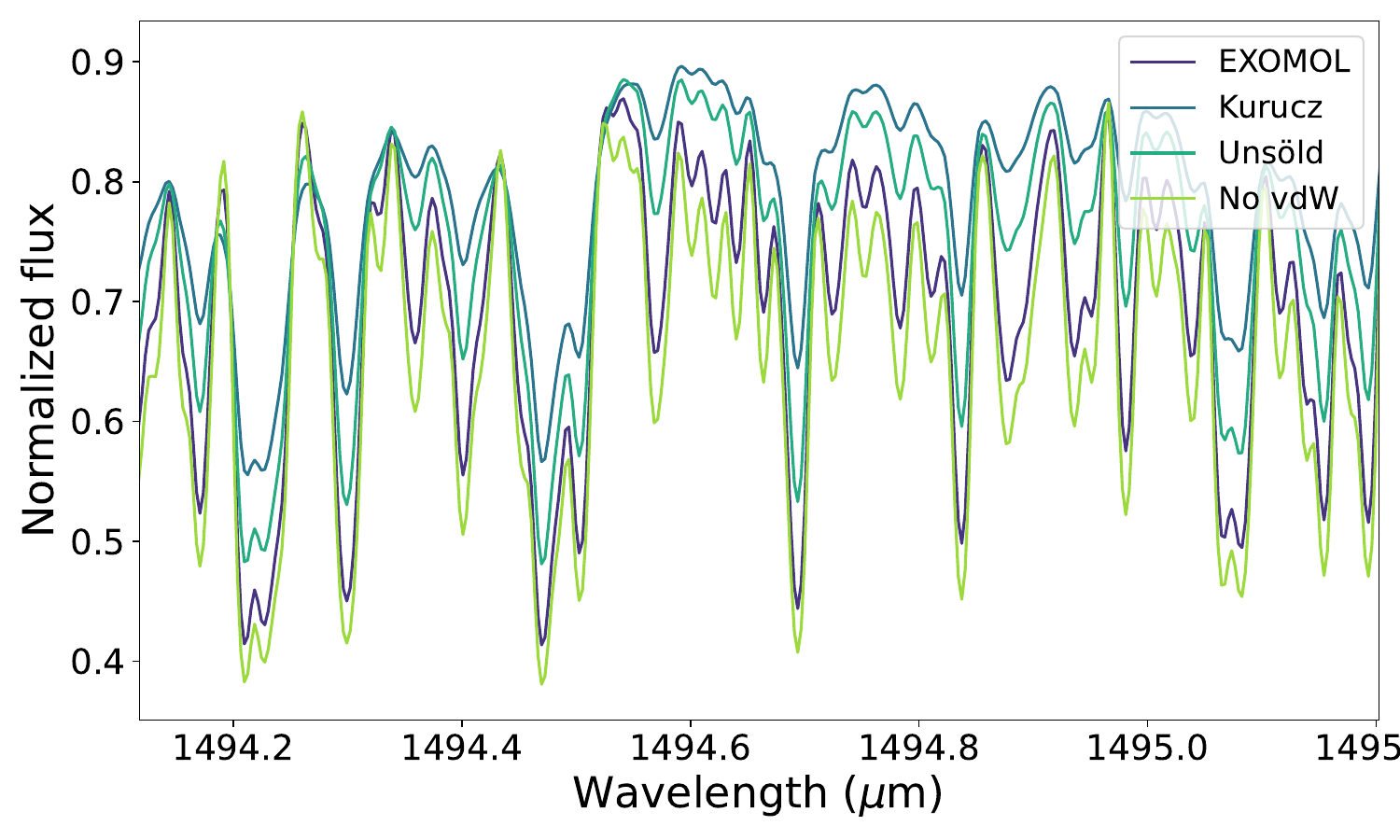}
    \caption{\textbf{Comparison of broadening implementations.} The normalized flux is shown on the y-axis and wavelength is given in microns on the x-axis. We show the impact of different broadening implementations at high resolution (R=500,000). Measured EXOMOL is shown in purple, Kurucz approximation in dark blue, Uns\"{o}ld approximation is shown in light blue, and no van der Waals broadening is shown in green. The EXOMOL broadening parameters are the most consistent with using no van der Waals broadening compared with the earlier approximations.}
    \label{fig:h2obroadening}
\end{figure*}

As an additional test, we compare different van der Waal broadening implementations for H$_2$O, where we exclude all other lines. First, we adopt the broadening parameters used within EXOMOL \citep{VORONIN_20102308,PETROVA_201650, Tennyson2016}. We use only the most abundant isotopologue. For each transition, we use the given broadening parameters. If no parameters are specified for a given transition, the line is not broadened. We then compare this broadening implementation over (a) no broadening, (b) the Uns\"{o}ld broadening approximation as outlined in \citet{Schweitzer_1996}, and (c) the Kurucz broadening implementation as described in \citet{Castelli_2005} with the same arbitrary broadening parameter as previously used in this work taken from Kurucz's line lists. In Figure \ref{fig:h2obroadening}, we show the impact of these broadening implementations. Using no van der Waals broadening is the most consistent with the EXOMOL broadening parameters derived for the water transitions compared with the previously considered approximations. This is consistent with our finding that using no van der Waals broadening finds the best agreement with the data.

\subsection{Alternative Explanations for the Reduced Strength of FeH Absorption}

Overestimation of pressure broadening in models is not the only possible explanation for the mismatch between observed and synthesized molecular features in the TRAPPIST-1 spectrum. TRAPPIST-1 is notoriously active, and magnetic activity can alter the structure of the photosphere. Frequent flares on TRAPPIST-1 can also lead to the formation of a hot chromosphere. The activity effects cannot be accounted for in the static 1D modeling presented in this study, but could modify the appearance of the spectrum.

Another possible explanation could be the formation of dust in the coolest regions of TRAPPIST-1 atmosphere \citep{Piaulet-Ghorayeb2025, Xuan2024}. For example, the condensation of iron into dust grains would reduce the amount of iron available for FeH formation and thus decrease the strength of the feature. Oxygen could also be incorporated into clouds, leaving less available  to form water, though TRAPPIST-1 may be too hot for silicate condensation \citep{Allard2012}.


\section{Summary and Discussion} \label{sec:summary}

With JWST, we are at the cusp of observing potential atmospheres of habitable-zone, terrestrial planets around M dwarf stars \citep{Espinoza2025, Glidden2025}. To access molecular features, we need to reach a precision $\sim$10 ppm \citep{Lustig-Yaeger2019}, which should be possible with the JWST noise floor \citep{Rustamkulov2022}. Strong features, such as those from CO$_2$, may be on the order of 50-200 ppm \citep{Lustig-Yaeger2019}. However, stellar contamination can be on the order of $\sim$200 ppm for a transiting Earth-twin planet around a TRAPPIST-1-like (M8V) star \citep{Rackham2018}. Accurate modeling of M dwarfs is the key bottleneck to accurately separate the stellar and planetary contributions of an observed spectrum. Here we showed that robust treatment of stellar pressure broadening is one critical part of our ability to interpret spectral data for signatures of terrestrial planet atmospheres.

We compared JWST NIRISS/SOSS data of active M dwarf TRAPPIST-1 to show the impact of proper van der Waals broadening approximations on the agreement between stellar models and data. We show that van der Waals broadening has been overestimated, leading to significant discrepancies between the models and data, in particular, around the $0.99~\micron$ FeH feature and the range between the large water absorption bands, which cause a peak in the pseudocontinuum around $1.7~\micron$. Our results showcase the need to carefully test each underlying assumption of stellar models to best exploit the gain in precision we have reached with JWST. Broadening is just one piece, and there remain many additional discrepancies between the models and data. In particular, 3D models are needed to accurately represent surface features \citep{Witzke2022, Smitha2025}. As a first step, modeling approximations inherited from hotter stars must be carefully evaluated before they can be used for cool M dwarfs.

\begin{acknowledgments}

We would like to thank the anonymous reviewer for their thoughtful and constructive comments, which strengthened the manuscript. ERC Synergy Grant: This work was founded by the European Research Council (ERC) under the European Union's Horizon 2020 research and innovation program (grant No. 101118581 - project REVEAL) and JWST GO \#3593 (PI: S. Seager). We thank Olivia Lim for the use of her reduced NIRSS/SOSS data of TRAPPIST-1 (doi:10.17909/qfgw-hv13). A.G. acknowledges the MIT Office of Research Computing and Data for providing high-performance computing resources that have contributed to the research results reported within this Letter. This work has made use of the VALD database, operated at Uppsala University, the Institute of Astronomy RAS in Moscow, and the University of Vienna.
\end{acknowledgments}

\vspace{5mm}
\facility{JWST NIRISS/SOSS}

\software{Astropy \citep{astropy:2013, astropy:2018, astropy:2022}, MPS-ATLAS \citep{Witzke2021}, POSEIDON \citep{MacDonald2023}, SpectRes \citep{Carnall2017}}

\bibliography{references}{}
\bibliographystyle{aasjournal}

\end{document}